# Concept Drift Adaptive Physical Event Detection for Social Media Streams

Abhijit Suprem[1], Aibek Musaev[2], and Calton Pu[1]

[1] Georgia Institute of Technology, GA 30332, USA
[2] University of Alabama, AL, 35487, USA
`asuprem@gatech.edu`

**Abstract.** Event detection has long been the domain of physical sensors operating in a static dataset assumption. The prevalence of social media and web access has led to the emergence of *social*, or *human* sensors who report on events globally. This warrants development of event detectors that can take advantage of the truly dense and high spatial and temporal resolution data provided by more than 3 billion social users. The phenomenon of *concept drift*, which causes terms and signals associated with a topic to change over time, renders static machine learning ineffective. Towards this end, we present an application for physical event detection on social sensors that improves traditional physical event detection with concept drift adaptation. Our approach continuously updates its machine learning classifiers automatically, without the need for human intervention. It integrates data from heterogeneous sources and is designed to handle weak-signal events (landslides, wildfires) with around ten posts per event in addition to large-signal events (hurricanes, earthquakes) with hundreds of thousands of posts per event. We demonstrate a landslide detector on our application that detects almost 350% more landslides compared to static approaches. Our application has high performance: using classifiers trained in 2014, achieving event detection accuracy of 0.988, compared to 0.762 for static approaches.

## 1 Introduction

The ubiquitous presence of web data and increase in users sharing information in social media has created a global network of *human reporters* who report on live events. Such human reporters can be considered as social sensors that provide information about live physical events around the globe [4-7]. Development of applications that can take advantage of social sensors to perform physical event detection are a clear next step. The primary challenge lies in the actual event detection: since social sensor data is a noisy text stream, machine learning models are required. Further, there is the phenomena of concept drift, where the distribution of real-world data changes with respect to time. This is especially apparent with text data. We provide an example with *landslide* detection on social media: the word *landslide* can refer to the disaster event or elections, among others. Since October and November are often election seasons in the United States, classification models that are tuned to ignore social media data with election related landslide keywords are more appropriate in this data window. In other months, the presence of election-related landslide tweets is scarce, which can cause increase false negatives in models overfit for election-related landslide tweets. This instance of changing data distributions is a form of concept drift.

So, social sensors constitute a challenge for the traditional approaches to text classification, which involve static machine learning classifiers that are never updated. Event detection systems for social sensors without concept drift adaptation face performance deterioration. We can consider Google Flu Trends (GFT) as an example; GFT was originally created to identify seasonal trends in the flu season [9]. However, the models did not incorporate changes in Google's own search data, causing increasing errors after release [9-11]. Our application addresses these challenges by incorporating concept drift adaptation. Additionally, we develop *automated* concept drift adaptation techniques to automatically generate training data for machine learning model updates. This is necessary due to the sheer volume of social media data – it is impractical to manually label the millions of social media posts per day. Our approach allows us to perform drift adaptation without human labelers, which significantly reduces training bottlenecks.

Specifically, we have the following contributions:

1. We present a drift-adaptive event detection application that performs physical event detection on social sensors using machine learning. We also show automated classifier updates for concept drift adaptation.
2. We develop a procedure to combine news articles and physical sensor data (e.g. rainfall data from NOAA and earthquake data from USGS) to perform automated training data generation for concept drift adaptation. Our application uses the low-latency, abundant social sensor data to perform physical event detection with machine learning classifiers, and the high-latency, scarce physical sensor data to tune and update classifiers.

We demonstrate our event detection application with landslide detection. We select landslides because they do not have dedicated physical sensors (in contrast to tsunamis or earthquakes); however, they cause large monetary and human losses each year. Landslides are a also what we call a *weak-signal* disaster: landslide-related social media data has significant noise in social media streams. Also, usage of *landslide* keywords to reference disasters is small compared to usage to reference irrelevant topics such as election landslides.

We compare our landslide detection application (LITMUS-adaptive) to the static approach in [2], which we call LITMUS-static. Our approach, LITMUS-adaptive, detects 350% more landslide events than LITMUS-static. We also evaluate our adaptive classifiers' accuracy compared to the static classifiers in LITMUS-static. We train classifiers with data in 2014, and compare performance of static and adaptive approaches in 2018. LITMUS-adaptive has f-score of 0.988 in 2018, compared to f-score of 0.762 in LITMUS-static, showcasing improvements our drift adaptive approach makes.

The rest of the paper is organized as follows: Section 2 covers related work. Section 3 covers data sources used in our application. Section 4 and 5 provide implementation details for our application. Section 6 and 7 evaluate our application quantitatively and qualitatively, respectively. Section 8 presents our conclusions.

## 2 Related Work

### 2.1 Physical Event Detection on Social Sensors

Earthquake detection using social sensors was initially proposed in [1]. There have also been attempts to develop physical event detectors for other types of disasters, including flooding [2], flu [3, 4], infectious diseases [5], and landslides [6, 7]. In most cases, the works focus on large-scale disasters or health crises, such as earthquakes, hurricanes [8], and influenza that can be easily verified and have abundant reputable data Our application is general purpose, as it can handle small-scale disaster such as landslides and large-scale disasters. The existing approaches also assume data without concept drift. However such assumptions, made in Google Flu Trends [9, 10] degrade in the long term.

### 2.2 Concept Drift Adaptation

Recent drift adaptation approaches evaluate their methods with synthetic data [11-14]. Such data is perturbed to include specific, known forms of drift. Several mechanisms have been developed for handling concept drift with numeric, sensor data.

**Windowing, or sliding windows,** is a common technique for adaptation. This approach uses multiple data memories, or windows of different lengths sliding over incoming data. Each window has an associated model. The SAM-KNN approach uses k-NN classifier to select window closest to a new data sample for classification [15]. Nested windows are considered in [16] to obtain training sets.

**Adaptive Random Forests** augment a random forest with a drift detector. Drift detection leads to forest pruning to remove tress that have poor performance. Pruned trees are replaced with new weak classifiers [17].

**Knowledge Maximized Ensemble (KME)** uses a combination of off-the-shelf and created drift detectors to recognize various forms of drift simultaneously. Models are updated when enough training data is collected and removed if they perform poorly [18].

Most methods approach concept drift with an eye towards detection and subsequent normalization. Updating or rebuilding a machine learning model facing drift involves two bottlenecks in the classification pipeline: data labeling and model training; of these, data labeling is the greater challenge due to its oracle requirements. Such wait-and-see models that perform corrections once errors have been detected entail periodic performance degradation before they are corrected with model up-dates; this may be infeasible in mission-critical applications. Active learning strategies counteract this bottleneck in part; the tradeoff is between highly accurate models and clustered, knowledge-agnostic representations that consider data on distance without subject matter expertise.

## 3 Data Sources

Our application, LITMUS-adaptive, combines physical sensor and news data, which have high-latency and are scarce, with social sensor data, which have low-latency and are abundant. The social sensor data is used for event detection through machine learning classifiers, while the physical sensors and news data are used to update machine learning classifiers.

### 3.1 Physical Sensors and News

1. **News**: News articles are downloaded from various online RSS feeds. Each source is described by the article link, the publish date, the article headline, and the publication name. Locations are extracted from the text using Named Entity Recognition. Publisher sources include international feeds from agencies (e.g. BBC, CNN, ABC, Reuters), as well as local news sources (some sample snippets are provided in Figure 1).
2. **Rainfall Reporting**: We download rainfall data from NOAA and earthquake data from USGS to perform to validate landslide detections.
3. **Landslide Predictions**: The National Oceanic and Atmospheric Administration (NOAA) and USGS provide landslide predictions in select locations where there is enough terrain and rainfall data. LITMUS uses this to perform localized landslide tracking and labeling.

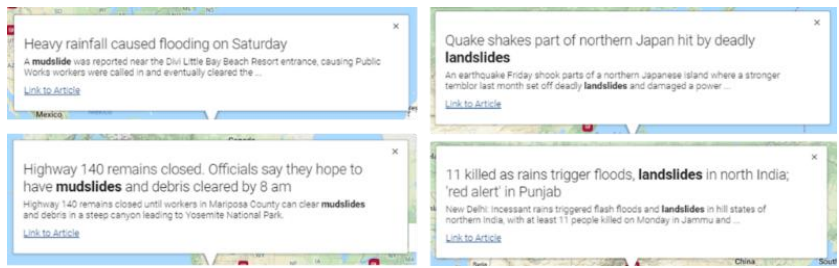

**Fig. 1.** Snippets of news articles about landslides. Each retrieved article is geolocated using NER to identify locations and indexed spatiotemporally.

### 3.2 Social Sensors

1. **Twitter**: a keyword streamer is used to download tweets continuously for Twitter. Keywords include the words 'landslide', 'mudslide', and 'rockslide' as well as their lemmas (some examples are provided below).
2. **Facebook**: a general keyword streamer is used to download public Facebook posts. Existing web crawlers are leveraged to improve retrieval efficiency

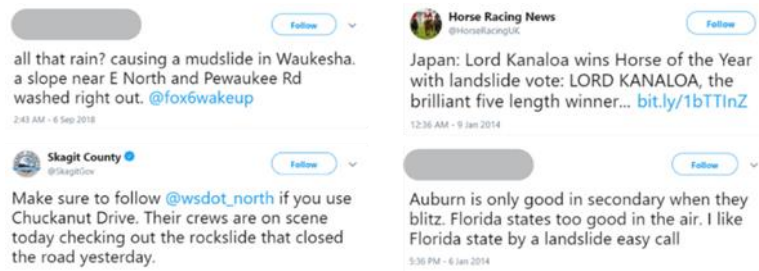

**Fig. 2.** Sample of raw tweets. The left side are relevant tweets. The right side are irrelevant tweets for landslide detection.

## 4 Approach

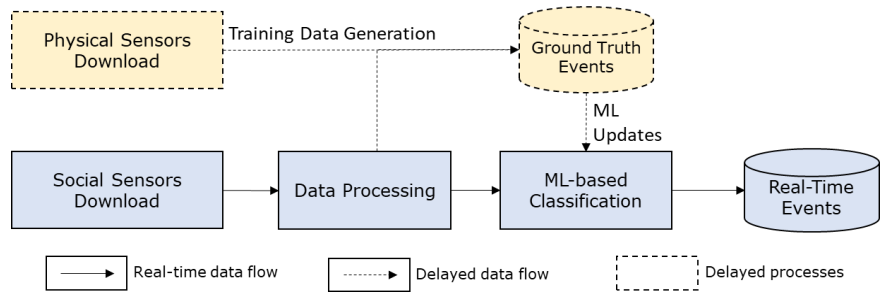

**Fig. 3.** Our application detects real-time events from social sensors. It remains drift-adaptive by integrating physical sensors with social sensors to continuously update machine learning classifiers.

The dataflow for our application is shown in Figure 3. We perform physical event detection by performing classification on social media data. We use binary classifiers that detect whether a given social sensor post is relevant to a given event or not (e.g. landslides). The latency between a physical event's occurrence and social sensor post about the event is significantly lower than latency with news reports and physical sensors, which often require expert confirmation. In contrast, social sensor data has low latency. However, it lacks the reputability of physical sensors and news reports. We rely on the social sensors for event detection, and physical sensors to continuously tune machine learning models.

Traditional approaches perform event detection under a static data assumption. Our contribution is in generating updated machine learning classifiers without any manual intervention so that our application can adapt to concept drift.

### 4.1 Social Sensor Download

Social sensor download operates in real-time. Our application has streaming endpoints for several short-text social media systems, such as Twitter and Facebook, with an extensible framework for integrating custom streaming sources. Social sensor downloads operate in a high-volume streaming setting. Each downloaded post object contains at least five fields: (i) the post content as a Unicode string, (ii) an array of named locations within the post (this is usually null and is filled during data processing step), (iii) timestamp of post, (iv) array of hyperlink content within the post, and (v) user-id or screen name of human reporter who created the post.

### 4.2 Physical Sensor and News Download

Physical sensors and news sources are dedicated physical, social, and web sensors providing event information with human annotations. In contrast to social sensors, these are trustworthy sources. We further distinguish physical sensors and news from social sensors: physical sensor and news data is highly structured and contains detailed event information. In our domain (landslides), most physical sensors and news sources provide geographical coordinates and time of landslide disaster. In most mission-critical applications, such physical sensor and news data appears long after an event takes place, once reputable sources have confirmed the event. Additionally, these sensors have lower volume.

For physical sensors, it is trivial to insert the physical event provided by the sensors into our ground truth database by extracting timestamp and location information. News articles provide topic tags that can be mined for an application's event; event reports (e.g. earthquake or large landslide report by USGS) provide detailed information about events, including locations, timestamps, event range, and event impact. We use Named Entity Recognition as well as location tags of news articles to extract location information to identify event location. These events are also stored in the ground truth event database as confirmed landslide events.

### 4.3 Social Sensor Data Processing

Social sensor data has low-context, which hinders location extraction and classification. Additionally, Named Entity Recognition (NER) often fails on short-text because there are too few words for the NER algorithms. We augment natural language extractors by sharing information between ground truth events and social sensor data processing. We provide an example with location extraction. Since social posts have few words, location extraction is not accurate on the short-text and often misses locations provided in a post's text content. So any location identified by NER is saved in memory for several days as a string. This string is used to augment location extraction with substring match; intuitively, if there is one social post about an event in a location, there could be others. Similarly, any ground-truth event locations are also added to the short-term memory to augment social sensor location extraction.

### 4.4 Automated Training Data Generation

We integrate social sensor data with physical sensors and news sources to automatically generate labeled data. During live operation, social sensor data is passed to the Machine Learning classifiers for event detection. Concept drift causes performance deterioration in these classifiers. So, our application performs model updates at regularly scheduled intervals. A model update requires labeled data, and a classifier's own labels cannot be used for updating itself. As noted before, it is also impractical to manually label the large volume of social sensor data (on Twitter alone, there are >500M tweets per day).

Our application matches historical social sensor data to ground truth events detected from physical sensors and news reports, which are highly trustworthy. Intuitively, social posts with landslide keywords that have similar space-time coordinates as ground truth events are very likely relevant to a real landslide (as opposed to irrelevant posts such as *election landslides*).

At the end of each data window (one-month windows in our landslide application), ground truth events of the window are stored as cells (coordinates of events are mapped to 2.5-min cell grids on the planet [7]). Social media data from the window is localized by time into 6-day bins. Note that in data processing step, NER augmentation only occurs *forward* in time, i.e. when a new location is added to the memory for substring matching, only subsequent social sensor posts are processed with the new location. So, during the automated training data generation stage, we have access to archived social sensor data.

**Table 1.** Automatically labeled data in each data window

| Data Window | Data Samples | Labeled |
|---|---|---|
| 2014-Training Data | 26,953 | 13028 |
| 2014-Test Data | 6464 | 3266 |
| July 2018 | 378 | 189 |
| August 2018 | 212 | 106 |
| September 2018 | 386 | 193 |
| October 2018 | 498 | 249 |
| November 2018 | 1770 | 885 |
| December 2018 | 446 | 223 |

We take advantage of this by re-processing data from the prior window. Locations extracted from each ground-truth event within $\pm3$ days are used as a substring filter to extract locations from social posts. We then perform automated labeling by matching a social media post's space-time coordinates to true-event location and time. Location matching is achieved with the 2.5-min cell grid superposition using strong supervision. This is in contrast to weak-supervision [21] as our supervisory labeling is domain-specific, instead of domain-agnostic. Table 1 below shows statistics about the generated training data (labeled data) in each window.

## 5 Event Detection with Machine Learning Classifiers

Our application uses machine learning classifiers to perform event detection. We employ a variety of statistical and deep learners, including SVMs, Logistic Regression, Decision Trees, and Neural Networks. Our drift adaptation consists of two complementary parts: **Classifier generation** (to create new ML models) and **Classifier updates** (to update older models with new data). We cover them below. We will first describe the classifier generation/update procedure. Then we will cover update schedules which govern when new classifiers are generated or updated.

### 5.1 Classifier generation

We define a window as the collection of social sensor data between two updates. At the end of a data window, training data is generated for the previous window using procedures in *Automated Training Data Generation*. The labeled samples are used to train new ML classifiers. Existing classifiers are copied, and the copies are updated with the new data. Both new and updated classifiers are saved to a database using key-value scheme. The classifiers function as values and the training data as the key. Currently, instead of storing the entire training data, we store the training data centroid as the key for a classifier.

### 5.2 Update Schedule

We support three types of classifier update schedules: *user-specified*, *detector-specified*, and *hybrid*, described below. These schedules allow for continuous classifier generation and updates to combat concept drift.

**User-Specified.** Users can set up an update schedule (daily, weekly, monthly, etc). The application tracks the internal time, and when an update is triggered, procedures in *Classifier Generation* are followed to create new classifiers and update existing ones.

**Detector-Specified.** Some classifiers types provide confidence values with predictions. Neural networks with softmax output layer provide class probabilities. Drift can be detected by tracking frequency of low confidence labels. For linear classifiers (including SVMs), higher density of data points close to the separating hyperplane over time can indicate signal drift. If drift time exceeds a threshold, procedures in *Classifier Generation* are followed to create new classifiers and update existing ones.

### 5.3 ML-Based Event Detection

Classifiers are retrieved using two approaches: (i) *recency* or (ii) *relevancy*. *Recency* performs lookup on most recently created classifiers. *Relevancy* performs k-NN (k nearest neighbors) search to find training data (using the stored data centroids) that is closest to prediction data; the k-closest data are then used to look up respective classifiers in the classifier database. These classifiers are used as an ensemble).

Our machine learning event detectors use multiple classifiers with votes to perform predictions, since ensembles perform better than lone classifiers. Our ensemble classifier supports several weight assignment options:

**Unweighted average.** The class labels predicted by each classifier in the ensemble (*0* for irrelevant and *1* for relevant physical event) are summed and averaged.

$Score \geq 0.5$ indicates majority of classifiers consider the input post as relevant to the physical event

**Weighted average.** Classifiers can be weighted by domain experts based on which algorithm they implement. Weak classifiers (random forests) can be given lower weights than better classifiers (SVMs, neural networks).

**Model-weighted.** We can determine classifier weights using their prior performance:

$$w_{M_i^t} = \frac{f_{M_i^{t-1}}}{\sum_a^n f_{M_n^{t_M}}}$$

where $M_i$ is classifier $i$, $w_{M_i^t}$ is the weight of classifier $i$ in data window $t$, and $f_{M_i^{t-1}}$ is the validation accuracy of $M_i$ on the testing data in the data window $t_M$ ($t_M$ is the window where the model was last trained).

## 6 Evaluation of Drift Adaptive Approach

We first demonstrate the need for concept drift adaptive event detection with evidence of concept drift in our data. As shown in Table 1, we have social sensor data from 2014 through 2018. Each labeled post's text is converted to a high-dimensional, numeric representation using word2vec [22]. The post vector is dimensionally reduced with tSNE. The tSNE-based reducer measures pairwise similarities between data points and shows the event characteristics separation between 2014 and 2018 data. Our real-world live data will continue to evolve over time.

Such drift in text data is difficult to predict due to the phenomena of lexical diffusion [23]. Current drift adaptation methods use synthetic data with bounded and predictable drift to evaluate methods; they do not focus on adapting to unbounded drift in live data.

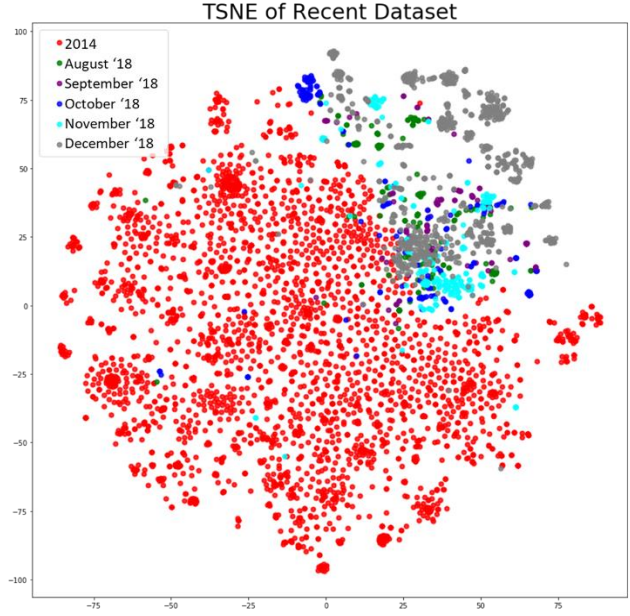

**Fig. 4.** TSNE of datasets from 2014 through 2018

We evaluate our drift adaptive event detectors with two approaches summarized in Table 2. In each window, the drift adaptive learner is provided with classifiers trained under each approach and tuned with optimal hyperparameters obtained using grid search.

**Table 2.** Summary of approaches

| Approach | Description | Training Data |
|---|---|---|
| N_RES (Static) | Non-drift resilient approach with static classifiers | 2014 Data |
| RES (Adaptive) | Drift resilient approach using generated training data for updates | 2014 Data – 2018 Data (Separated by windows) |

We compare performance of each approach in Table 2 over subsequent windows in our data. As we show in Figure 5, ensembles with resilience (adaptive approach) outperform non-resilient counterparts (static approach) throughout. RES under both statistical and deep learners maintains high f-score across multiple years. N_RES has significantly higher variance in performance, and is generally poor at adapting to the live data's drift.

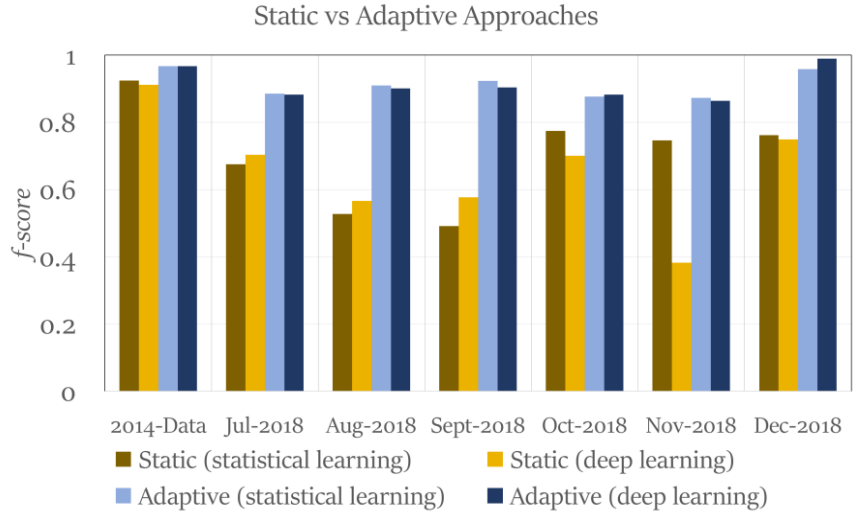

**Fig. 5.** The Adaptive methods significantly outperform static counterparts in the 2018 data.

N_RES classifiers face deterioration across all metrics without access to generated training data to update their parameters (Figure 6).

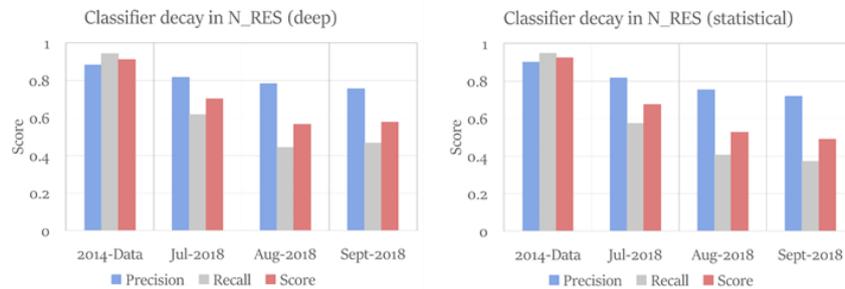

**Fig. 6.** Decay is apparent in non-resilient classifiers. The f-score is high during the offline window, but degrades without access to generated training data.

## 7  Landslide Detection: Results

We built our application with physical event detection in mind; evaluation is performed with landslide detection. In the previous section, we validated our drift adaptive approach. Here, we compare our application – LITMUS-adaptive, to LITMUS-static, the traditional approach. Figure 7 shows the raw event comparisons between LITMUS-static and LITMUS-adaptive. LITMUS-adaptive outperforms LITMUS-static, and over time, the share of events detected only by LITMUS-adaptive increases (Figure 8). We see that by December 2018, LITMUS-adaptive detects 71% of events, compared to LITMUS-static's 28% of events, an increase by 350%. We show global coverage in Figure 9.

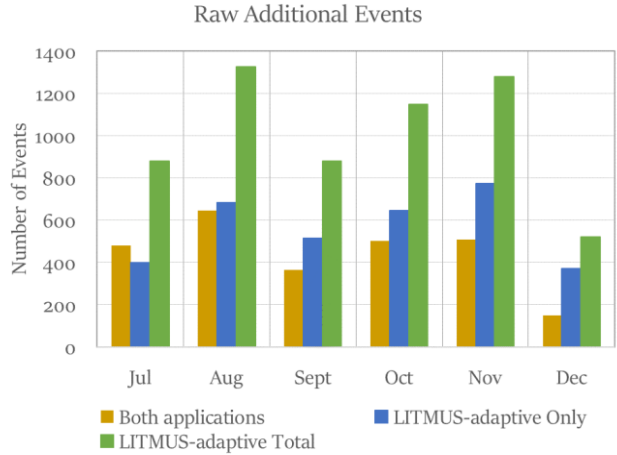

**Fig. 7.** Raw additional events comparison between LITMUS-adaptive and LITMUS-static. *Both applications* are events detected by both LITMUS-static and LITMUS-adaptive. Every event detected by LITMUS-static was also detected by LITMUS-adaptive. In addition, LITMUS-adaptive also detects several additional events (*LITMUS-adaptive only*). The sum of the two are shown in *LITMUS-adaptive Total*

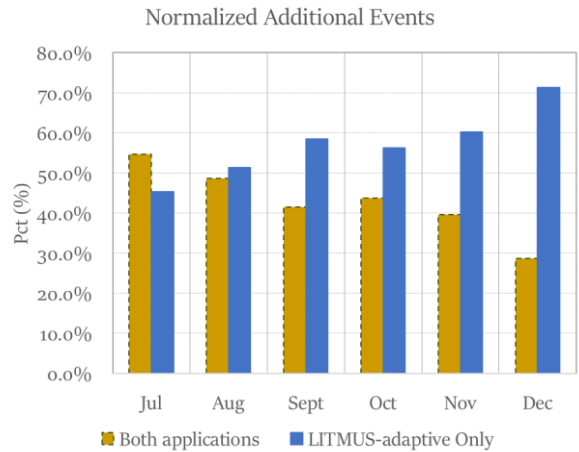

**Fig. 8.** Events normalized as fractions of total. Over data windows, fraction of data missed by LITMUS-static increases as its classifiers deteriorate. However, with continuous updates, LITMUS-adaptive can adapt to drift and maintain higher accuracy and thus, better event detection.

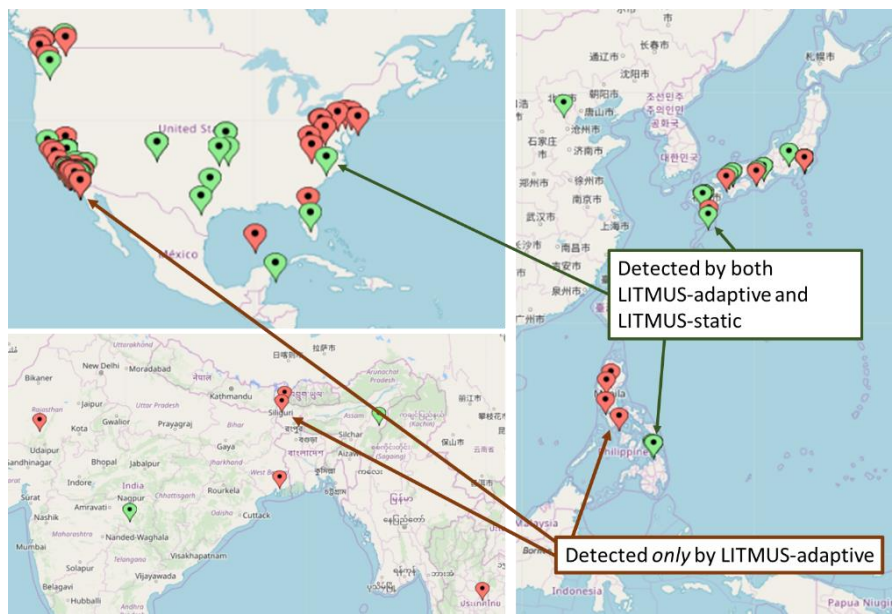

**Fig. 9.** Our application has global coverage. We are able to detect events across the globe, and as mentioned, every event in LITMUS-static is also detected in LITMUS-adaptive, along with hundreds of additional events LITMUS-static misses (Figure 7 and Figure 8).

## 8 Conclusions

We proposed a physical event detector for social sensor data that remains resilient to concept drift. Our approach combines social sensors with physical sensors and news data to perform *continuous learning* to maintain model currency with the data distribution. Our application's drift adaptation takes advantage of human annotations in existing reputable sources (physical sensors and news data) to augment and generate training data. This removes the need for humans to perform manual labeling, significantly reducing cost and labeling bottlenecks.

Our application uses an ML-based event processing classifiers that continuously adapt to changes in live data. We believe the application and the methods presented in this paper can be useful for a variety of social-sensor based physical event detection. We demonstrated a disaster detection application that is designed for landslide detection. Our application, LITMUS-adaptive, improves upon static approaches such as LITMUS-static. LITMUS-adaptive adapts to changing event characteristics in social sources and detects almost 350% more landslide events than LITMUS-static. Moreover, LITMUS-adaptive achieves f-score of 0.988 by December 2018, compared to f-score of 0.762 for static approaches.

## 9 Acknowledgement

This research has been partially funded by National Science Foundation by CISE's SAVI/RCN (1402266, 1550379), CNS (1421561), CRISP (1541074), SaTC (1564097) programs, an REU supplement (1545173), and gifts, grants, or contracts from Fujitsu, HP, Intel, and Georgia Tech Foundation through the John P. Imlay, Jr. Chair endowment. Any opinions, findings, and conclusions or recommendations expressed in this material are those of the author(s) and do not necessarily reflect the views of the National Science Foundation or other funding agencies and companies mentioned above.